\documentclass[sigconf,screen]{acmart}
\usepackage[utf8]{inputenc}
\usepackage{amsmath,amssymb,amsfonts}
\usepackage{listings}
\usepackage{xcolor}
\usepackage{hyperref}

\definecolor{gray}{RGB}{96,96,96}
\definecolor{gold}{RGB}{128,128,0}
\definecolor{red}{RGB}{127 0 0}
\definecolor{orange}{RGB}{224,64,0}
\definecolor{brown}{RGB}{128,64,0}
\definecolor{navy}{RGB}{0,0,160}
\definecolor{turqoise}{RGB}{0,127,160}
\definecolor{green}{RGB}{0,127,0}
\definecolor{lila}{RGB}{127,0,160}

\newcommand{\HIDE}[1]{}
\newcommand{\IVXR}{{\sf iv4XR}}
\newcommand{\unblocked}[1]{\lfloor #1 \rfloor}

\renewcommand\footnotetextcopyrightpermission[1]{}
\acmConference {11th Int. Workshop on Automating Test Case Design, Selection, and Evaluation (ATEST)}{2020}{}

\title{Navigation and Exploration in 3D-Game Automated Play Testing}
\titlenote{{\Large\bf PREPRINT}. This is the author's version of the work. It is posted here for your personal use. Not for redistribution. The definitive version was published in the Proceedings of ACM SIGSOFT ATEST-2020, DOI: \url{https://doi.org/10.1145/3412452.3423570}
}
\author{I.S.W.B. Prasetya}
\orcid{0000−0002−3421−4635}
\affiliation{\institution{Utrecht University}}
\email{s.w.b.prasetya@uu.nl}
\author{Maurin Voshol}
\affiliation{\institution{Utrecht University}}
\email{j.m.voshol@uu.nl}
\author{Tom Tanis}
\affiliation{\institution{Utrecht University}}
\email{t.a.tanis@students.uu.nl}
\author{Adam Smits}
\affiliation{\institution{Utrecht University}}
\email{a.smits2@students.uu.nl}
\author{Bram Smit}
\affiliation{\institution{Utrecht University}}
\email{b.smit@students.uu.nl}
\author{Jacco van Mourik}
\affiliation{\institution{Utrecht University}}
\email{j.vanmourik@students.uu.nl}
\author{Menno Klunder} 
\affiliation{\institution{Utrecht University}}
\email{m.c.klunder@students.uu.nl}
\author{Frank Hoogmoed}
\affiliation{\institution{Utrecht University}}
\email{f.d.hoogmoed@students.uu.nl}
\author{Stijn Hinlopen}
\affiliation{\institution{Utrecht University}} 
\email{f.a.hinlopen@students.uu.nl}
\author{August van Casteren}
\affiliation{\institution{Utrecht University}}
\email{a.p.b.vancasteren@students.uu.nl}
\author{Jesse van de Berg}
\affiliation{\institution{Utrecht University}}
\email{j.s.vandenberg3@students.uu.nl}
\author{Naraenda G.W.Y. Prasetya}
\affiliation{\institution{Utrecht University}}
\email{n.g.w.y.prasetya@uu.nl}
\author{Samira Shirzadehhajimahmood}
\affiliation{\institution{Utrecht University}}
\email{s.shirzadehhajimahmood@uu.nl}
\author{Saba Gholizadeh Ansari}
\orcid{0000-0002-7135-5605}
\affiliation{\institution{Utrecht University}}
\email{s.gholizadehansari@uu.nl}

\renewcommand{\shortauthors}{\IVXR}

\begin{document}

\begin{abstract} 


To enable automated software testing, the ability to automatically navigate to a state of interest and to explore all, or at least sufficient number of, instances of such a state is fundamental. When testing a computer game the problem has an extra dimension, namely the virtual world where the game is played on. This world often plays a dominant role in constraining which logical states are reachable, and how to reach them. 
So, any automated testing algorithm for computer games will inevitably need a layer that deals with navigation on a virtual world. 
Unlike e.g. navigating through the GUI of a typical web-based application, 
navigating over a virtual world is much more challenging.
This paper discusses how concepts from geometry and graph-based path finding can be applied in the context of game testing to solve the problem of automated navigation and exploration.
As a proof of concept, the paper also briefly discusses the implementation of the proposed approach.

\end{abstract}



\maketitle

\section{Introduction}

The computer games industry has been around for quite a long time, 50 years, since the first game console was brought to the market in early 70s.  The industry has become huge, with estimated revenue of over 120 billions USD in 2019 (to grow to 200 billions in 2023) \cite{Newzoo20}.
As it is now, computer games have simply become an inseparable part of our contemporary culture.

Quite paradoxically though, the technology supporting
its quality assurance lags quite far behind that of other types of software industries, such as services, web-based applications, or even mobile applications.
The common practice to test a computer game is still by 'play-testing' it, where human users are deployed to play the game to run a bunch of test scenarios and subsequently report their finding (e.g. if they notice visual, functional, or performance issues).
The process is mostly manual and therefore expensive. Technology for automated testing, except at the very rudimentary level, is practically absent.
Perhaps it is because the research community, or even the industry itself, do not really consider computer games to be business critical, and hence have less drive to invest in automated testing. 
Regardless the reason, the effect is the same: the high cost of play testing, and its slow process (we cannot expect to get feedback in less than an hour), slow down the development process and hamper companies' ability to quickly market new games. Especially for small companies and startups this is a major disadvantage.

To automate play testing one can think of applying well known automated testing techniques such as Model-based Testing (MBT) \cite{utting2012taxonomy} or Search-based Testing (SBT) \cite{mcminn2004search}. However, directly trying to apply them to computer games is likely to be ineffective due to the very fine grained level of interactivity of computer games. 
When the player controls his/her in-game character by moving his/her mouse, the game samples the mouse at the rate of 60-100 times/second. So, every update will only move the character a tiny distance in the game world. While this improves the sense of realism, at this level of details the game state space would be extremely large to be handled by the aforementioned techniques. An abstraction layer needs thus to be introduced to make the state space tractable for available testing techniques such as MBT and SBT.

Imagine the task of verifying $N$ levers in some game level, that each correctly opens the right in-game doors. To do this by automated play testing, imagine we introduce a software agent that automatically plays the corresponding in-game character that a human tester would play.
Abstractly, it is the same task as verifying the return value of $N$ functions. At this level, the problem is easily tractable by even random testing. In the play testing setup however, the human play tester would first need to navigate to each lever, pull it, and then he/she would have to navigate to relevant doors to check that only the right one is open. Since the test agent is intended to simulate a human tester, it also has to do the same. 
To simply teleport to the levers might give an incorrect verdict, e.g. if one of the levers turns out to be unreachable from the player's starting location\footnote{The issue is analogous to checking correctness by partial versus total correctness. }. 
Notice that navigation over the game's virtual world is thus part of the problem. If the test agent does not know how to navigate to a lever, it cannot test it either. 
Another way to look at this is that equipping test agents with automated world navigation and exploration will allow them to test the game at the more functional level, and hence enabling the integration of techniques such as MBT and SBT.

This paper will show how concepts from geometry and automated path finding can be applied in the context of game testing to provide the aforementioned auto-navigation and exploration skills to test agents. 
While the subjects were much studied in other contexts, e.g. robot control \cite{kwek1997simple,panaite1999exploring}, we have not seen it discussed in the context of software testing, so  we hope that this paper also provides useful insight for the software testing research community in developing techniques and better tools for testing computer games. As motivated before, the industry can really use such a push.
As a proof of concept, we ourselves have implemented the approach as part of the \IVXR\ project\footnote{\url{https://github.com/iv4xr-project}}, which is a project to provide an open source agent-based automated testing framework to test extended reality based systems (which include computer games)\footnote{
   While \IVXR\ tries to be generic, one should keep in mind that every game is in many ways unique as it has a unique world with its own ontology, along with a unique way to interact with it. No general testing framework, including \IVXR, can be plugged into an arbitrary game without some integration effort. However, in return the developers will then get access to \IVXR's features such as agent-based programming, reasoning-based AI,
and auto-navigation and exploration as this paper will discuss
(and more in the future as we work on adding more AI capabilities).} \cite{iv4xrICST20,AplibEmas}.
A demo-project is also available\footnote{\url{https://github.com/iv4xr-project/iv4xrDemo}}, demonstrating features of \IVXR's automation on a demo game called Lab Recruits\footnote{\url{https://github.com/iv4xr-project/labrecruits}}.

The paper is organized as follows. Section \ref{sec.setup} describes the basic testing setup that we will assume. 
Section \ref{sec.nav} discusses the problem of auto-navigation in the context of automated play testing, and how it can be solved.
Section \ref{fig:exploration} discusses auto-exploration, which is needed if the locations of the target in-game entities that need verification are not known upfront, or have changed due to some design decisions.
Section \ref{sec.cov} discusses several concepts of navigation-related test coverage.
Section \ref{sec.relatedwork} discusses some related work, and finally Section \ref{sec.concl} concludes and mentions some future work.

\section{Automated Play Testing: Basic Setup} \label{sec.setup}

\begin{figure}[htp]
    \centering
    \includegraphics[width=6cm]{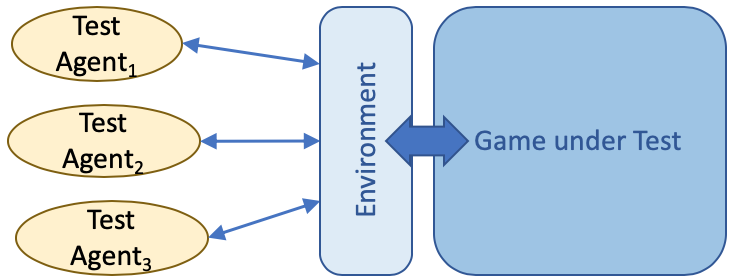}
    \caption{A simple agent-based setup of game testing.}
    \label{fig:setup}
\end{figure}

Figure \ref{fig:setup} shows a simple setup of agent-based game testing using \IVXR. The idea behind the setup is however general, namely to use agents which are programmed to play and test a given game, and hence they can be used to replace human play testers. Multiple agents can be deployed to simulate multiple play testers, if the game has a multiplayer mode.
In \IVXR, the agents would be programmed in Java; the framework furthermore allows reasoning and goal-based strategies to be added to implement complex plays. 

For simplicity, in this paper we will only consider a single test agent setup.
We will assume a typical execution model of an agent-based system \cite{MeyerAgentTech2008,AgentDesignPat2016}, \IVXR\ agents also follow this model:
an agent runs in cycles, interacting with an 'environment'. In our case, this environment is a game under test, or some interface to this game under test.
At every cycle, the agent {\em observes} its environment, {\em deliberates}, and then {\em sends a command }to the environment to try to steer it towards some goal state\footnote{We can furthermore distinguish between  synchronous and asynchronous setups. In the first the agent controls how much the game can progress between cycles, whereas in the latter the control is absent. For testing, the first setup is preferable as it makes test runs more repeatable. This aspect is however beyond the scope of this paper.}. 
The agent controls an in-game player character. The commands it sends to the game are typically primitive commands, such as moving the controlled character a small distance, or to make it interacts with another in-game entity within in its interaction range.

Importantly, we will assume that when the agent requests observation, the game under test will send back a {\bf structurally represented observation} (e.g. listing relevant in-game entities and their states) rather than simply sending an image. Having structural observation allows much more accurate inspection of the game state, rather than having to depend on image recognition.

Since the chosen testing approach is 'play testing', it is important that the agent behaves as an actual human tester. This means that {\bf the control and observation by the agent should abide to the same constraints that apply to human players}. E.g. the agent cannot just teleport to a location, nor can it see through a solid wall.


\section{Automated Navigation for Play Testing} \label{sec.nav}

Imagine a testing task to verify the state of a certain in-game entity at a certain location in the target virtual world. Obviously, it would be helpful if the test agent can automatically navigate to the goal location.
An algorithm to do this may already be implemented in the game under test itself, e.g. to guide the movement of in-game enemy entities. The test agent can in principle reuse it.
There are however several considerations that may prevent such reuse: the implementing module may not be accessible to an external agent, and besides that, a test agent may need test-specific features that are not present in the game's native auto-navigation. 
For this reason, our \IVXR\ Framework offers its own navigation and exploration modules which are better suited to support testing related tasks. But, before we discuss the more testing-related aspects, let us first discuss how to do auto-navigation in general.

\begin{figure}[htp]
    \centering
    \includegraphics[width=3.8cm, height=3.8cm]{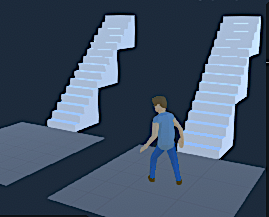}
    \includegraphics[width=4cm]{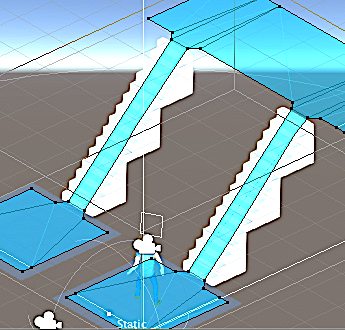}

    \caption{A simple world (left) created with the Unity. The world has two ground-floor rooms reachable from each other
    through stairs via another floor (not shown). The picture on the right shows the navigation mesh (blue) of this world. The mesh is generated by the game engine of Unity.}
    \label{fig:navigation-mesh}
\end{figure}

To go from location $A$ to $B$ in some virtual world, simply taking the straight line between them will not work in most situations as there are often obstacles between them. Directly trying to find a path by quantifying over all possible paths does not work either, because most virtual worlds form in principle a continuous space, and hence the number possible paths is infinite.
A typical solution used in many games, e.g. those using the Unity game engine, is to first divide the navigable surface of the target 3D world into a so-called {\em navigation mesh} \cite{millington2019AI}, consisting of a finite number of connected polygons (also called {\em faces}) as shown in the example in
Figure \ref{fig:navigation-mesh}. 
Finding a path in the world can now be reduced to the problem of finding a path over the mesh.
This navigation mesh is typically static, so if the game under test already has it, we just need to extend it a bit so that it exports this mesh. As we will see below, having this mesh will allow the test agent to do its own auto-navigation.

\begin{figure}[htp]
    \centering
    \includegraphics[width=8cm]{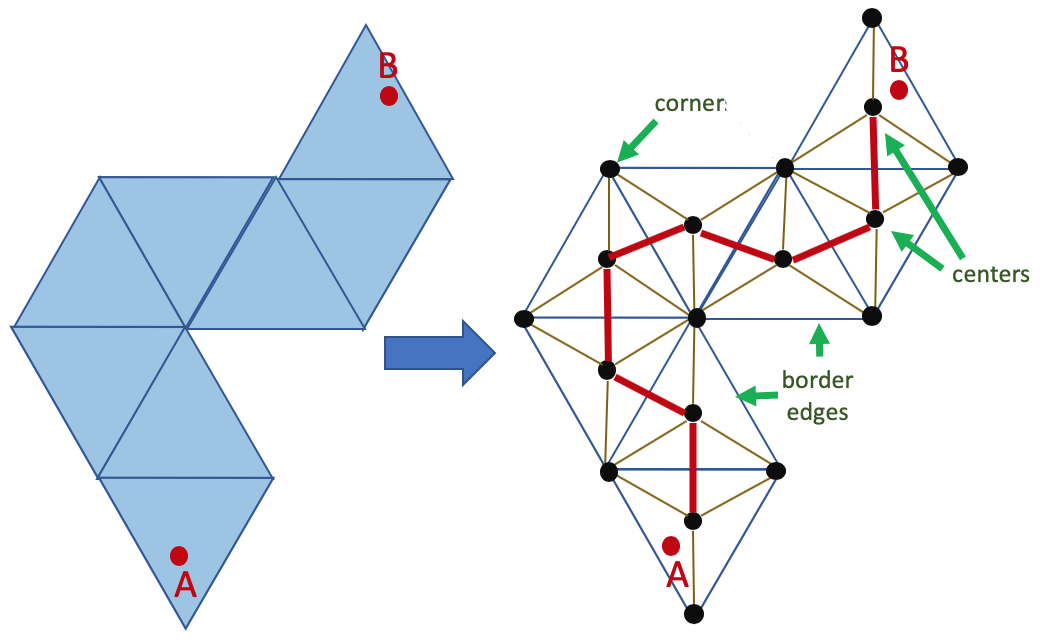}
    \caption{From a navigation mesh (left), we construct a navgraph (right), which is graph over location-vertices (black). We can now do navigation over this finite graph.}
    \label{fig:navgraph}
\end{figure}

To actually do navigation the agent converts the mesh to a {\em navigation graph}, also called {\em navgraph}, $G = (V,E)$ where $V$ is a set of vertices, each represents a location on a face in the mesh and $E$ is the set of edges between them, such that when $(a,b)\in E$ the straight line between them is physically navigable in the game's virtual world and is clear of any (static) obstacle. 
Any path in $G$ is thus guaranteed to be physically navigable. Such a $G$ can be constructed from a navigation mesh by taking the corners of the faces as the vertices, connected by the corresponding edges in the faces.
Figure \ref{fig:navgraph} shows an example.
Additionally, since a human player tends to walk in the middle, e.g. when moving through a corridor, we will also add the center points of the faces to $V$. We also connect centers of neighboring faces (faces with an edge in common), as indicated by
 the red lines in Figure \ref{fig:navgraph}.

Let $G.{\sf pathfinder}$ be a chosen graph-based path finding algorithm, such as A* \cite{hart1968formal, millington2019AI} or Dijkstra \cite{dijkstra1959note}. 
Given two vertices $a,b\in V$, $G.{\sf  pathfinder}(a,b)$ will return a path over the edges of $G$ that goes from $a$ to $b$, provided $b$ is indeed reachable from $a$.

So, to auto-navigate from $A$ to $B$, e.g. as in Figure \ref{fig:navgraph},
the agent first searches the set $V$ for vertices $a$ and $b$ which are closest to respectively $A$ and $B$. Then, invoking $G.{\sf pathfinder}(a,b)$ will give the agent the path to go from $a$ to $b$,

\subsection{Dynamic obstacles and hazards} 

Many games have dynamic obstacles, whose state may change at the runtime, influencing which parts of the world they block. An example is an in-game fence or door. When closed, it is an obstacle, but otherwise it is not. Obviously, the test agent should also know how to deal with such an obstacle. 

Many game worlds also have hazardous areas, such as a lava field
in the a navigable area, though walking over it is hazardous for the agent, e.g. it might die. In most cases we would want the test agent to steer away from such a field, but not always. There may be a game level where crossing over a lava field is necessary to get to some goal state. To flexibly deal with hazardous areas we can therefore treat them as dynamic obstacles, whose state the test agent can flip from blocking to non-blocking depending on whether it wants to avoid them or not. 

\begin{figure}[htp]
    \centering
    \includegraphics[width=5cm]{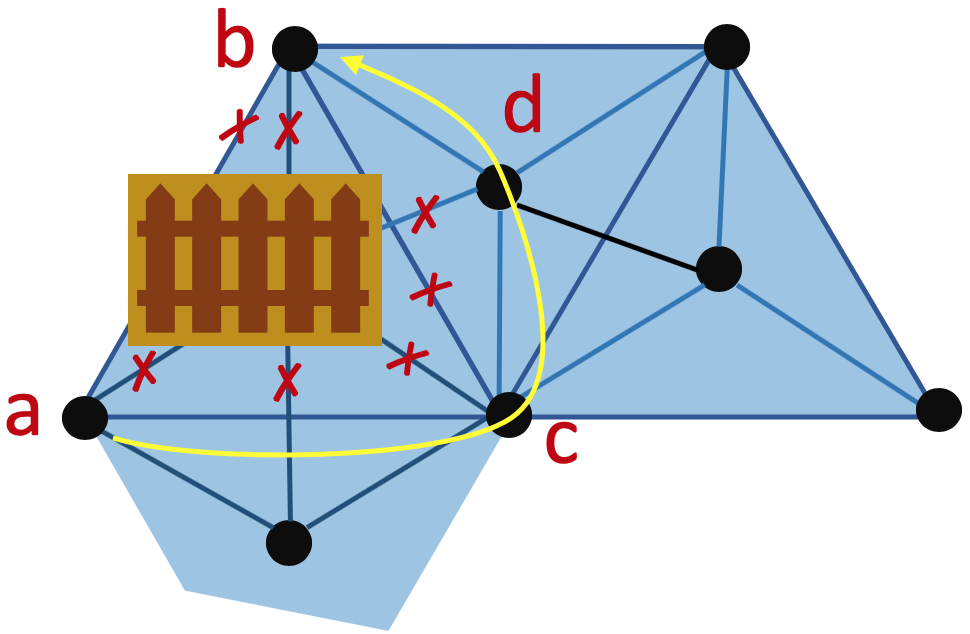}
    \caption{In the above example the agent wants to go from $a$ to $b$, but as it is about to do that the fence changes its state from open to close.}
    \label{fig:obstacle}
\end{figure}

To do dynamic obstacle avoidance, we extend the edges $e$ in the $G$ to have a state $e.{\sf clear}$, which is true if the edge is clear, and false if it is blocked by an obstacle. 
When the agent observe that the state of an obstacle $O$ has changed, and furthermore we assume that the observation includes the location and dimension of $O$, the agent would know which edges in $E$ are affected by the change, namely the edges whose lines intersect with $O$.
Figure \ref{fig:obstacle} shows an example. The blue area represents the a navigable surface in the game under test. 
A part of its navgraph is overlayed on this surface, with dark circles indicate its vertices connected with edges.
Imagine the test agent wants to go from $a$ to $b$, and then a fence becomes closed. The edge $(a,b)$ represents the straight line route from $a$ to $b$. Unfortunately, this line intersects with the fence, so the agent cannot take this route to go to $b$. Likewise, all edges marked with a red-cross are also blocked, as long as the fence stays closed. Fortunately $b$ is still reachable, e.g. through the yellow colored path, which the test agent can follow to get to $b$.

Let $\unblocked{G}$ be the subgraph of $G$ consisting of only clear (unblocked) edges, and the vertices they connect. To avoid obstacles, the test agent should not invoke $G.{\sf pathfinder}$, but rather: $\unblocked{G}.{\sf pathfinder}(a,b)$.
Obviously, for this to work the test agent will need to track the states of dynamic obstacles.
This is another reason why we cannot always rely on the game's native auto-navigation, as it may not be designed to deal with dynamic obstacles, e.g. when it is used to steer enemy units, the route may be as such that these enemy units will never encounter dynamic obstacles anyway.

\subsection{Finer auto-navigation control}

For testing purposes we may want the auto-navigation to take different routes than 'normal', e.g. the ones that the game's native auto-navigation would come up with.
E.g. the native auto-navigation may insist on a path that avoid hazard, while a test agent may occasionally want to take a path through some hazard.
The solution discussed in the previous section can coerce a test agent to consider a hazardous area as navigable, but this is not always sufficient to actually force the agent to navigate through it.
This implies that a test agent needs to have a finer control on the kind of path that its path finding algorithm should look for. 

For each edge $e\in E$ let's define $e.{\sf cost}$ to be the cost of traversing the edge. Its natural (default) cost is just the geometric distance between its end points (the length of the straight lines between these end points). The cost of a path is the sum of the cost of its edges. Path finding algorithms such as A* and Dijkstra are {\em optimal} as they will return the best path (one with the least cost) to the asked destination.

To influence $G.{\sf pathfinder}$ the agent can define a set $G.{\sf prefs}$ of preferred vertices. The cost of an edge $(a,a')$ is increased over its natural cost if $a' \not\in G.{\sf prefs}$. This will encourage an optimal path finder to find a path through preferred vertices\footnote{An alternative would be to {\bf decrease} the cost of going to preferred vertices. However, note that this may not work well if we use A* as the path finder. A* can find the best path very fast, but it requires so-called heuristic cost (estimated cost) of going between any two vertices $(a,b)$ to be provided. This heuristic cost should ideally be an under-estimation of the real cost, or else A* might not return the best path. A good and easy measure to be used as the heuristic distance is the geometric distance between $a$ and $b$. Now, if we adjust the real cost of some edges, to become less than the geometric distance between their end points, this may break the assumption on the heuristic cost defined as previously mentioned.}.

For example, marking center vertices (in Figure \ref{fig:navgraph} these are vertices connected by red lines) as preferred will encourage the test agent to choose a path through the center of corridors. In contrast, marking 'border vertices' will encourage the test agent to walk along the walls, e.g. suitable if we want to randomly check if the walls are indeed solid (the character cannot pass through them).
A vertex is a {\em border vertex} if it belongs to a border edge. A {\em border edge} is an edge that is not shared by two faces.

\section{Less Positional-dependent Testing: Exploration}
\label{sec.exploration}
The approach described before enables a test agent to find an in-game entity, assuming its physical location is known. However, relying on locations to find entities that require checking is not a robust testing approach,  
as such a test would break if the game designer decides to move the entities to a different location; something that happens often during the game development.
A better approach, inspired by   good practices in GUI testing \cite{Screenster20,Robula}, is to assign unique ids to in-game entities. If an entity $x$ with id $i$ cannot be found in its old location, the test agent can decide to search the virtual world, until it sees an entity $x'$ with $x'.{\sf id} = i$. Since the $i$ is unique, $x'$  must thus be $x$ itself. 

The above will require the agent to systematically search the virtual world. 
This can be made simpler if the exported navigation mesh is fine grained enough with respect to the agent's visibility range ({\bf mesh granularity assumption}). That is, the size of the faces should be as such that the distance of each face's corners to the face's center should be less that the agent's visibility range.
This guarantees that every point in (the navigable part of) the virtual world is visible from some vertex in the navgraph. Note that if the mesh exported by the game is too coarse grained, we can always split the faces to get them to the desired granularity.

An exploration algorithm that visits all vertices in $G$, such as Depth First Search (DFS), is thus guaranteed to find $x$ (actually, exploring all centers of the faces is sufficient). However, things get more complicated due to dynamic obstacles. 
DFS may not work because a dynamic obstacle may close the agent's backtracking path.
An alternative is to use a DFS-like exploration algorithm over a directed graph \cite{kwek1997simple}. Because the graph is directed, such an algorithm does not assume that it can just backtrack to the previous vertex, but instead explicitly looks for a route to the backtrack vertex.
Both algorithms assume the agent to have the visibility range of 0. That is, an agent can only discover one unexplored vertex (which is then marked as 'explored') at a time.
However, most in-game character has a visibility range $>0$.
This means, when an agent traverses to an unexplored vertex $b$, it might also see more unexplored vertices, e.g. $c$ and $d$. We can exploit this.

\begin{figure}[htp]
    \centering
    \includegraphics[width=5cm]{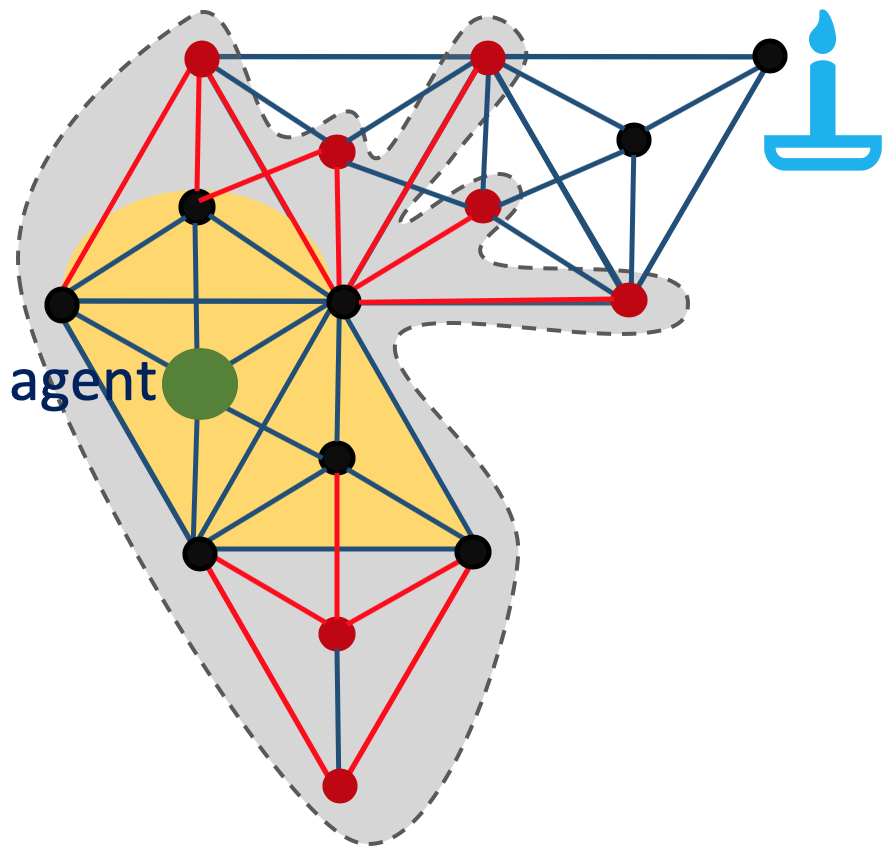}
    \caption{In this example the agent wants to navigate to the candle, but it does not know its location. So, it explores the navgraph to find it. The yellow region indicates the part of the game world it has seen so far. The red vertices form the current exploration frontier. The agent chooses one of the frontier vertex to navigate next.}
    \label{fig:exploration}
\end{figure}

We therefore propose the following frontier-based exploration.
To illustrate the algorithm imagine the navgraph is as in Figure \ref{fig:exploration}, and the test agent is searching the entity $x$ indicated by the candle (e.g. it wants to check that the candle is not lighted).

\begin{enumerate}
    \item Maintain a set $S$ of the vertices in $G$ that is marked as 'explored'; initially only the currently visible vertices are in $S$.
    
    \item \label{startLoop} In Figure \ref{fig:exploration} imagine now that the agent is at the green node and the yellow area is the area explored so far. So, all the vertices in this area are in $S$.
    
    Calculate the set $F$ of {\em frontier vertices}.
    These are vertices in $V/S$ with one edge that connect them to some vertex in $S$, and are moreover reachable through unblocked paths from the agent current position (indicated green in 
    Figure \ref{fig:exploration}).
    The frontier vertices are marked red in Figure \ref{fig:exploration}.
    
    \item If $F$ is empty the exploration is done (there is no more unseen and reachable vertex to explore to).
    
    Otherwise choose one frontier vertex $f$ from $F$ that is preferred, e.g. such that $f$ is the closest to the agent's current position, or else simply choose randomly. 
    Then calculate a path to $f$,
    and navigate to it.
    
    \item Whenever an agent sees a vertex that is not yet in $S$, it is added to $S$. If the agent also sees the goal entity $x$, the exploration is done.
    
    \item Repeat from step \ref{startLoop}.
\end{enumerate}

Keep selecting the closest frontier to explore will yield an algorithm that behaves much like DFS. When the agent sees no more frontier in its close vicinity, it will go to the next one a bit further away, which is comparable to DFS's backtracking. However, the next frontier to explore does not have to be in the agent's backtracking path. While DFS will get into a problem if the backtracking path becomes closed, our algorithm can still progress as long as there is a frontier left.

No pure exploration algorithm can however deal with a dynamic obstacle that persistently cuts off the access to some vertices unless the agent manages to somehow flip the obstacle's state. This requires some reasoning to be incorporated into the agent, which furthermore can be very game specific. This is outside the scope of this paper.

\section{Coverage}\label{sec.cov}

'Coverage' is a metric of the sufficiency of a test. Since we use navgraph, this naturally suggests the use of graph-based coverage concepts \cite{ammann2017introduction} such as vertex coverage. However, not all of them would make sense (e.g. trying to cover all non-cycling paths would quickly become unfeasible), and which one to use may depend on the testing task at hand. Below we mention few examples:

\begin{itemize}
    \item The task is to verify the state of an in-game entity $x$, but the agent cannot find $x$. If all vertices in the navgraph are explored, this implies that $x$ does not exists, and therefore the test verdict is a 'fail'. If not all vertices are explored (or at least, all center vertices), the verdict is 'inconclusive'.
    
    \item The task is to verify the state of {\em all} entities of type $T$. Suppose the agent manages to find $N$ of them, and they are all in the correct state. We would want to have full vertex coverage, or else the verdict would still be inconclusive.
    
    \item The task is to check that all walls are 'solid'. That is, the agent cannot walk through them. We would want to have coverage over all border edges.
    
    \item The task is to verify that the game has no cheat spot, e.g. where the player cannot be shot at. A typical place for such a spot is a corner. So, we would want to cover all corners.
    They can be identified as border vertices, whose outgoing border edges form an angle of less than e.g. $150^\circ$ (there is one marked example in Figure \ref{fig:navgraph}).
    
    \item The task is to verify that some entity $x$ in a room $R$ can only be reached by crossing a lava field. If checking all possible (non-cyclic) paths is not feasible, we can introduce different path types. E.g. paths that only go through the center vertices in $R$, the path that clockwisely follows the border edges, and the path that counter-clockwisely follows the border edges. We can then insist on covering all these types of paths.
    
\end{itemize}

\section{Implementation}

\begin{figure}[htp]
    \centering
    \includegraphics[width=7cm]{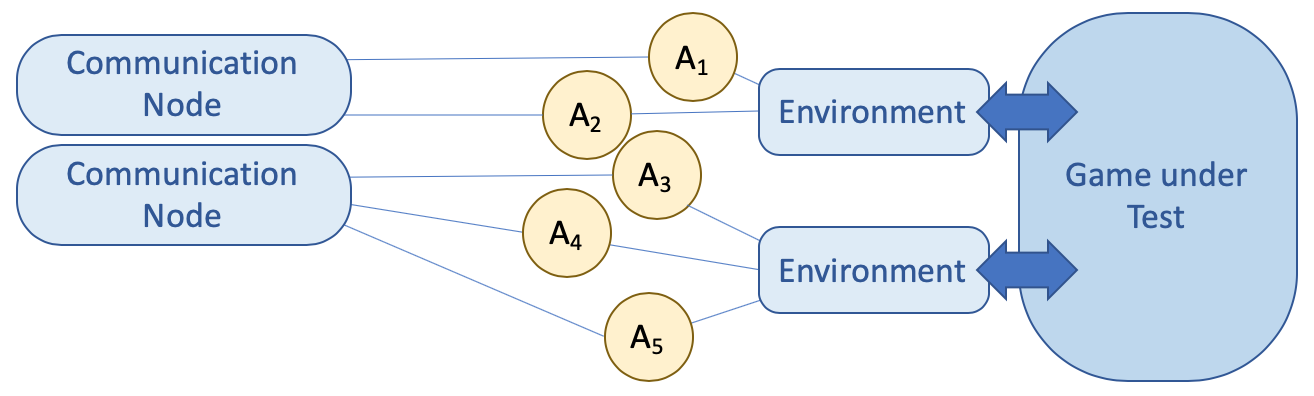}
    \caption{A simple agent-based setup of game testing.}
    \label{fig:setup2}
\end{figure}

As a proof of concept, 
the auto-navigation and exploration discussed in Sections \ref{sec.nav} and \ref{sec.exploration} have been implemented as part of the automation in the agent-based testing framework \IVXR.
Figure \ref{fig:setup2} shows a more elaborate setup than in Figure \ref{fig:setup} for using \IVXR\ to test a computer game. Multiple test agents can be deployed. From an agent's perspective, it interacts with an 'Environment', which actually is a Java interface that handles the communication with the game under test.
An agent can register to a communication node, which will allow it to send and receive messages to/from other agents that register to the same communication node, thus enabling them to coordinate their actions.

The code snippets below shows a simple program of a test agent in \IVXR:

\begin{lstlisting}[mathescape=true,
  numbers=left, 
  numberstyle=\tiny,
  frame=leftline,
  xleftmargin=5mm]
SEQ(entityIsInteracted($button_1$),
    doorIsInRange($door_1$),
    entityInvariantChecked(testAgent,
      $door_1$, 
      (Entity x) $\rightarrow$ x.getBooleanProperty("isOpen")))
\end{lstlisting}	  

Agents are programmed in Java, in a goal-based style. The program above says that the agent has three {\em goals}: first (line 1) to interact with an in-game entity $\sf button_1$, then (line 2) to get $door_1$ in its visual range, and then (lines 3-5) to verify that $door_1$ is open. Solving a goal will require actions, which are programmed in so-called {\em tactics}. Each goal should have a tactic intended to solve it (not shown in the above snippet). For example, to interact with $button_1$ the agent first needs to be standing close to the button, which in turns requires it to first {\em navigate} to the button. The tactic that programs this navigation looks as shown below:

\begin{lstlisting}[mathescape=true,frame=leftline,xleftmargin=5mm]
FIRSTof(navigateTo(entityId), 
        explore(),
        ABORT())
\end{lstlisting}

A tactic is executed in an implicit loop. Recall that an agent executes in cycles.
At each cycle, the tactic of the agent's current goal is checked if it is enabled. If it is, it will be executed {\em for one cycle}. Else, the agent will do nothing, hoping that the game will change its state at the next cycle, which may enable the tactic, or until it runs out of budget and declares the current goal as failed.

In the above example, the tactic is to choose the first of its subtactics that is enabled on the current agent's state. The tactic ${\sf navigateTo}(x)$ implements auto-navigation discussed in Section \ref{sec.nav}, and will steer the agent to the location of entity $x$, if its location is known. Else the tactic is not enabled.
The tactic $\sf explore$ implements the auto-exploration from Section \ref{sec.exploration}. So, if the location of $x$ is unknown to the agent, it will do exploration. This tactic is enabled as long as there are frontier vertices to go to.
Note that full exploration is not always needed. Once the agent sees $x$, its position becomes known, and hence the tactic ${\sf navigateTo}(x)$ becomes enabled, and will be the one that will be chosen in the next cycle thanks to the $\sf FIRSTof$ operator.
If none of the above two tactics are enabled (implying the agent has searched, but cannot find $x$), we have $\sf ABORT$, that is always enabled, but will cause the current goal to be declared as fail.

For more on programming \IVXR, see \cite{AplibEmas} and the provided documentation of the framework\footnote{\url{https://github.com/iv4xr-project}}

\section{Related Work} \label{sec.relatedwork}

There are tools such as the Unity Test Framework\footnote{\url{https://unity.com/}} and GameDriver\footnote{\url{https://www.gamedriver.io/}} that allow game testing tasks to be scripted, hence they can be executed repeatedly by a computer rather than manually.
This works well for games with {\em coarse grained} interactions where every interaction corresponds to a functional behavior. Chess and most text-based adventure games are examples of such games. More generally, games that have no movable agent are coarse grained.
In contrast, games that are played on continuous 2D or 3D virtual worlds are usually {\em fine grained}. The scripting approach does not work well on such games, as scripting more complex testing tasks becomes increasingly tedious and error prone.
Some tools like GameDriver and MAuto \cite{Mauto19} can record a play and replay it as a test. 
While this removes the need to script the test, record and replay tests are unfortunately also fragile. They break when the game designers change the layout of a game level, or move the positions of in-game entities.

Examples of advanced automation can be found in the tools Icarus \cite{pfau2017automated} and Prowler \cite{Prowler}. Both use learning algoritms to learn how to do a given testing task. Once this is learned, the task can be executed repeatedly without manual intervention. Icarus' case study shows that the approach works for a coarse grained game. Prowler's example is Minecraft, which is a 3D, fine grained game, but only simple testing tasks were attempted. 
Compared to the above mentioned approaches, our \IVXR\ Framework \cite{AplibEmas} offers an approach that lies in between. As in the scripting approach, developers have to program the tests. However, \IVXR\ comes with various automation support such as reasoning and tactical programming, allowing tests to be programmed at a much higher level to handle fine grained games.
Auto-navigation and exploration are essential parts of \IVXR\ automation.
It is not a push-button technology like Icarus and Prowler, though on the other hand, being able to program the test agents makes \IVXR\ a versatile approach, and its insistence to use structured observation makes it highly accurate in its assessment. 

Learning as in Icarus and Prowler is not the only way we can obtain automation in testing. 
Model based testing (MBT)
and search based testing (SBT) 
can also give us automation.
MBT allows tests to be generated from a behavioral model of the system under test (e.g. in the form of an FSM).
MBT has been widely studied in other types of software systems, e.g. libraries, services, web applications, and mobile applications. 
The application of MBT for game testing has not been well studied though. Among the few work we could find was that of Iftikhar et al \cite{iftikhar2015automated}.
The case study is the Super Mario Brother, which is a fine grained game, thus implying that MBT can be applied to a fine grained game. However, the studied behavior is very limited and it does not involve much navigation. Also, in the standard MBT setup, the used model would describe the full behavior of the system under test. We should keep in mind that such a complete model for a game would be very large and too costly if it is to be manually crafted.

As a case for SBT, people have also studied the use of e.g. genetic algorithms to train an agent to play games, e.g. a game similar to Super Mario \cite{baldominos2015learning}. Although 'playing' is not literally the same as testing, it is conceivable that the same approach can be used to search an interaction sequence (a test case) that would automate a given testing task. 
Although the  game is fine grained,  the number of possible interactions is limited (just 22 in \cite{baldominos2015learning}, e.g. moving left, right, up, down, shooting, etc), which makes the search space tractable compared to e.g. Minecraft where the character can move in any direction in $360^{\circ}$ and hence its set of possible actions is infinite. For such a game, auto-navigation as discussed in this paper would be essential, basically to provide an abstraction that reduce the search space to become a finite space again.


 



\section{Conclusion and Future Work}
\label{sec.concl}

Ability to do auto-navigation and exploration is crucial for automated play testing of computer games. Without this ability, the more functional automation will not be possible. In this paper we have discussed the problems for games that are played in a continuous 2D or 3D virtual world, and how known concepts and algorithms from geometry and path finding can be applied to solve them.
Although many games do have built-in auto navigation algorithm, we cannot always reuse them for testing purposes, as the later need more refined control on the generated paths, e.g. to specifically explore the borders, or to avoid dynamic obstacles.
The most flexible solution would be to have a dedicated navigation and exploration library for testing purposes, such as the one provided by \IVXR, or for the game developers to invest in developing a common one for both for testing and in-game purposes. 

The presence of dynamic obstacles is prevalent in many games. A pure navigation algorithm can only deal with them up to a certain extent. A more powerful way to deal with them would be by incorporating reasoning on how to remove the obstacles when this becomes necessary. Unfortunately, such reasoning is likely to be quite game specific. 
But perhaps there may be common patterns/heuristics that work on similar games. This is future work.
This paper is also limited in discussing surface navigation (though the surface might be 3D). Testing games that allow full 3D movement (e.g. when the player drives a spaceship that can move freely in all directions, so not just on a surface) would need a different kind of navigation. This is also future work.


\begin{acks}
We thank Utrecht University Projectbureau for their support for this work. We also thank Raja Lala and Vedran Kasalica from Utrecht University for their support and assistance.

This work is supported by the \grantsponsor{H2020SponsorIDblabla}{EU ICT-2018-3 H2020 Programme}, grant nr. \grantnum{H2020SponsorIDblabla}{856716}.
\end{acks}

\bibliographystyle{ACM-Reference-Format}
\bibliography{mybib}    
\end{document}